\documentclass[twocolumn,aps,floatfix,superscriptaddress,prb,nobibnotes,longbibliography]{revtex4-1}

\usepackage{graphicx}
\usepackage{amsmath,amsthm,amssymb,amsfonts}

\usepackage[USenglish]{babel}
\usepackage{color}

\usepackage[colorlinks=true,linkcolor=blue,citecolor=blue]{hyperref}

\begin{document}
\title{Induced pairing of fermionic impurities in a one-dimensional strongly correlated Bose gas}

\author{Michael Pasek}
\altaffiliation[Present address: ]{Institut f\"ur Theoretische Physik, Goethe-Universit\"at, 60438 Frankfurt am Main, Germany}
\email{pasek@th.physik.uni-frankfurt.de}
\affiliation{Universit\' e de Paris, Laboratoire Mat\' eriaux et Ph\' enom\`enes Quantiques, CNRS, F-75013, Paris, France}
\affiliation{Laboratoire Kastler Brossel, UPMC-Sorbonne Universit\'es, CNRS, ENS-PSL Research University, Coll\`{e}ge de France, 4 Place Jussieu, 75005 Paris, France}

\author{Giuliano Orso}
\email{giuliano.orso@univ-paris-diderot.fr}
\affiliation{Universit\' e de Paris, Laboratoire Mat\' eriaux et Ph\' enom\`enes Quantiques, CNRS, F-75013, Paris, France}

\begin{abstract}
We investigate numerically the problem of few (one, two)  noninteracting spin$-1/2$ fermions in a shallow harmonic trap 
coupled via contact repulsive interactions to a uniform one-dimensional bath of lattice bosons, described by the Bose-Hubbard model. 
Through extensive density-matrix renormalization group calculations, we extract the binding energy and the effective mass of  
 quasiparticles, including dressed impurities (polarons) and their two-body bound states (bipolarons), emerging from the effective non-local Casimir interaction between the impurities.
We show that the mixture exhibits rather different pairing behaviors depending on the singlet \textit{vs}.~triplet spin state configurations of the two fermions.
For opposite spin states, bipolarons are found for  any finite value of the impurity-bath coupling. In particular, in the strong coupling regime 
 their binding energy reduces to that of a single polaron, provided the boson-boson repulsion is not too weak.
For equal spin states, we show that bipolarons emerge only beyond a critical strength of the Bose-Fermi interaction and 
their effective mass grows rapidly approaching the strong coupling regime.
\end{abstract}


\maketitle

\section{Introduction}
\label{sec:intro}
The recent experimental realization of mixtures of Bose and Fermi superfluids in dilute cold atomic gases~\cite{ferrier2014,delehaye2015,onofrio2016} has allowed to probe the complex many-body physics of coupled interacting spin-1/2 Fermi gases and superfluid Bose gases in a variety of parameter regimes, with the possibility of controlling the lattice geometry with tailored potentials and the strength of interactions using Feshbach resonances~\cite{bloch2008}. 
Previous theoretical works on interacting Bose-Fermi mixtures have  uncovered a variety of different ground-state phases~\cite{bijlsma2000,heiselberg2000,efremov2002,viverit2002,matera2003,enss2009,titvinidze2009,anders2012,bukov2014,bilitewski2015,Christensen:PRL2015,Levinsen:PRL2015,wu2016,midtgaard2016,kinnunen2018,Yoshida:PRX2018,Pierce:PRL2019}. 
One-dimensional Bose-Fermi mixtures, which show peculiar properties owing to the enhanced role of quantum fluctuations~\cite{cazalilla2011}, have also received a sustained interest~\cite{das2003,cazalilla2003,mathey2004,imambekov2006,pollet2006,mathey2007,sengupta2007,rizzi2008,mering2008,marchetti2009,orignac2010,fang2011,danshita2013,dekharghani2017,reichert2017,nielsen2018,siegl2018,huber2019inmedium,singh2019enhanced}. 

A  mobile fermionic impurity interacting with a surrounding bath of bosons is dressed by the collective excitations in the Bose gas, leading to the formation of a quasiparticle, the Bose polaron, with an enhanced effective mass~\cite{landau1948,froehlich1954,feynman1955}.
The Bose polaron problem for a single impurity, stemming originally from the physics of electron-phonon interactions~\cite{mahan2000}, has been extensively studied in recent years with ultra-cold atoms in one dimension~\cite{bruderer2007,schecter2012a,schecter2012b,casteels2012,bonart2012,bonart2013,massel2013,peotta2013,dutta2013,kantian2014,yin2015,dehkharghani2015,grusdt2015,schecter2016,petkovic2016,pastukhov2017,volosniev2017,parisi2017,grusdt2017a,mistakidis2019a}, and above~\cite{tempere2009,privitera2010,ardila2015,shchadilova2016,grusdt2017,grusdt2018,ardila2019}.
Recent experiments on impurities in a one-dimensional Bose gas~\cite{palzer2009,catani2012,fukuhara2013,scelle2013} and Bose-Einstein condensates in three dimensions~\cite{hu2016,jorgensen2016,rentrop2016,yan2019} have also been performed,
with successful measurements of the polaron self-energy, which can be extracted from spectroscopy measurements~\cite{jorgensen2016}, or the effective polaronic mass, from monitoring dipole oscillations of the impurity in a harmonic trap~\cite{catani2012}. 

The natural extension of these works on the Bose polaron problem for a single impurity is the study of few interacting impurities coupled to a common bath. 
This longstanding problem, first addressed in $^3$He -$^4$He mixtures~\cite{baym1967}, 
 has recently been studied theoretically in one-dimensional~\cite{dehkharghani2018,sarkar2018,schmidt2019,mistakidis2019b,mistakidis2019c} and three-dimensional~\cite{naidon2018,camacho2018a,camacho2018b}
cold atomic gases. Static and mobile impurities in one-dimensional quantum fluids are of particular interest because of the emergence of an effective 
attractive Casimir interaction between impurities due to the exchange of phononic excitations in the bath~\cite{recati2005,waechter2007,schecter2014,pavlov2018,reichert2018,reichert2019}.
Hence, two impurities can form a new bosonic bound state, the bipolaron, a quasiparticle state that has been studied in the context of high-$T_c$ superconductivity~\cite{alexandrov1994,waldram1996,devreese2009}.

In this work, we investigate the polaron problem for one and two 
fermionic impurities interacting with a one-dimensional Bose gas in a tight optical lattice described by the Bose-Hubbard model. Based on the density-matrix renormalization group (DMRG)~\cite{schollwoeck2011,dolfi2014}
 method, we compute the ground-state energy of the mixture in the presence of a shallow harmonic potential for the impurities
 as a function of the trapping frequency. This allows us to extract the binding energy and effective mass of both polarons and bipolarons, with great accuracy. For bipolarons, we show that these quantities exhibit distinct properties depending on the spin configuration, singlet or triplet, of the two constituent fermions.
 Our numerical approach goes beyond mean-field and variational calculations, and provides a competing alternative to quantum Monte Carlo methods~\cite{parisi2017,grusdt2017a} to probe the Bose polaron problem in one-dimensional systems.

Although purely induced pairing of trapped bosonic impurities coupled to a one-dimensional bath has been recently addressed theoretically~\cite{dehkharghani2018},
 the underlying pairing mechanism was essentially due to mean-field effects of the impurities on a trapped ideal Bose gas, and not due to  the exchange of phononic modes, which were absent. Moreover, no pairing between fermionic impurities was observed in this work.

The article is organized as follows. In Sec.~\ref{sec:model} we present the microscopic model Hamiltonian used to describe the Bose-Fermi mixture.
In Sec.~\ref{sec:single} we start by introducing our theoretical approach to compute the binding energy and effective mass of
a single polaron. We then present our numerical results based on the DMRG method and compare them with analytical predictions from Bogoliubov theory, in the regime of weak Bose-Fermi coupling.
In Sec.~\ref{sec:two} we extend this approach to bipolarons and present numerical results for the binding energy and effective mass, distinguishing the cases of two impurities with equal and opposite spins.
  
\section{Model Hamiltonian}
\label{sec:model}
We describe the one-dimensional Bose-Fermi mixture by a  lattice Hamiltonian consisting of three parts, 
$ \hat{H} = \hat{H}_b +\hat{H}_f + \hat{H}_{bf} $. The first term corresponds to the Bose-Hubbard model 
\begin{equation}
 \hat{H}_b = -t_b \sum_{\langle i, j\rangle} \hat{b}_{i}^\dagger \hat{b}_j  +\frac{U_b}{2} \sum_{i} \hat{n}_{ib} ( \hat{n}_{ib} -1 ),
 \end{equation}
 describing bosons hopping between neighboring sites with tunneling rate $t_b$ and subject to on-site repulsive interactions of strength $U_{b}>0$. 
  
Importantly, we assume in this work that the spin-$1/2$ fermions have no \emph{direct} intra-species interaction, while effective interactions between fermions
will be generated dynamically via the exchange of bosonic density fluctuations.
We also consider that fermions are trapped at the center of the chain 
by an harmonic potential. 
In the absence of coupling to the bosons, they obey the Hamiltonian
 \begin{equation}\label{HamF}
 \hat{H}_f =  - t_f \sum_{\langle i,j\rangle, \sigma} \hat{c}^\dag_{i \sigma} \hat{c}^{\phantom\dag}_{j\sigma}  
+ \sum_{i,\sigma}   \frac{1}{2 }m^*\omega^2 \left(i- \frac{L}{2}\right)^2  \hat{n}_{i\sigma}, 
 \end{equation}
 where  $\sigma=\uparrow,\downarrow$ accounts for the two spin states, $t_f$  is the fermion hopping rate,  $m^*=1/(2 t_f)$ is the bare fermionic mass in the lattice, $\omega$ is the trapping frequency and $L$ is the size of the chain. Here both $\hbar$ and the lattice period have been set to unity.  
 
Finally, boson and fermion densities are coupled by the following Hubbard-like term 
 \begin{equation}
 \hat{H}_{bf}= U_{bf}  \sum_{i,\sigma} \hat{n}_{ib} \hat{n}_{i\sigma},
\end{equation}
where $U_{bf}$ is the strength of the Bose-Fermi interaction, that we assume to be always repulsive, $U_{bf}>0$.

In this work, we assume that the boson density $n_b=N_b/L$ is finite,  $N_b=\sum_i\langle \hat{n}_{ib}\rangle$ being the total number of bosons in the system. 
The strength of boson-boson interactions in the Bose gas can then be characterized via the  dimensionless Lieb-Liniger parameter
  $\gamma= U_b / (2 n_b t_b)$~\cite{cazalilla2011,peotta2013}. 
In contrast,  the total number of fermions per spin component, $N_\sigma=\sum_i\langle \hat{n}_{i\sigma}\rangle$, is chosen 
finite and small, corresponding to the picture of few (one, two) fermionic impurities moving in a bath of  correlated bosons. 
For simplicity,  in the following we restrict to equal hopping amplitudes of  the fermionic and bosonic components  and fix the energy scale by setting $t_f=t_b =1$. 

 The impurity trapping potential in Eq.~(\ref{HamF}) allows us to extract the binding energy and effective mass of polarons 
 and bipolarons in a uniform system by carrying out DMRG calculations of the ground-state energy of the mixture
 for different values of the trap frequency and performing a scaling analysis, as discussed  in Secs.~\ref{sec:single} and \ref{sec:two}. 
The use of DMRG to extract the polaron mass through a finite size scaling analysis was first investigated for the one-dimensional Holstein model~\cite{jeckelmann1998}. 
Since our DMRG method requires open boundary conditions, we find that, in the absence of the trapping potential $(\omega=0)$, the mobile impurity in its ground-state binds at the edge of the chain, where the density of the bath is lower than its bulk value~\cite{dehkharghani2015,dehkharghani2018,reichert2019b}.
To avoid such strong finite-size effects, we restrict our calculations to finite $\omega$ values that keep the impurities in the middle of the chain.

\section{Single impurity}
\label{sec:single}
We first consider the properties of a single polaron, obtained by immerging a single fermionic impurity in the bosonic bath.
For definiteness, we choose the impurity to have spin up, i.e.~$N_\uparrow=1$, $N_\downarrow=0$.

\subsection{Trap scaling approach for polarons}
Let us assume, for the time being, that there is no trapping potential, $\omega=0$, and that the length of the chain is infinite, so that 
the lattice Hamiltonian  of the coupled system is translationally invariant.

Let  $E_q^\uparrow$ be the  lowest energy  of the mixture for a given momentum $q$ of the polaron. 
For small $q$, this energy can be Taylor-expanded as  
\begin{equation}\label{poladisp}
E_q^\uparrow \simeq E_0^\uparrow+\frac{q^2}{2m_p^*}+A^\uparrow q^4,
\end{equation}
where $E_0^\uparrow $ represents the absolute ground state energy, $m_p^*$ represents the polaron effective mass and $A^\uparrow$ is a numerical constant accounting for 
anharmonic terms in the polaron energy dispersion.  
 The binding energy $\mu_p$ of the polaron is defined as 
\begin{equation}\label{mu_p}
\mu_p=E_0^\uparrow-(E_\mathrm{gs}^{b}-2),
\end{equation} 
where $E_\mathrm{gs}^{b}$ is the ground-state energy of the interacting Bose gas in the absence of impurities and the last term corresponds to  the ground state energy
of a free fermionic impurity with energy dispersion relation $\epsilon_q^f = -2\cos q$.
In the absence of Bose-Fermi coupling, we have $E_q^\uparrow=E_\mathrm{gs}^{b}+\epsilon_q^f $, 
implying that the binding energy of the polaron vanishes, i.e.~$\mu_p=0$. Since $\epsilon_q^f \simeq  -2+q^2-q^4/12$,  
the polaron effective mass reduces to the bare mass of the impurity, $m_p^*=1/2=m^*$, as expected, and $A^\uparrow=-1/12$. 

\begin{figure}
     \includegraphics[width=0.95\columnwidth]{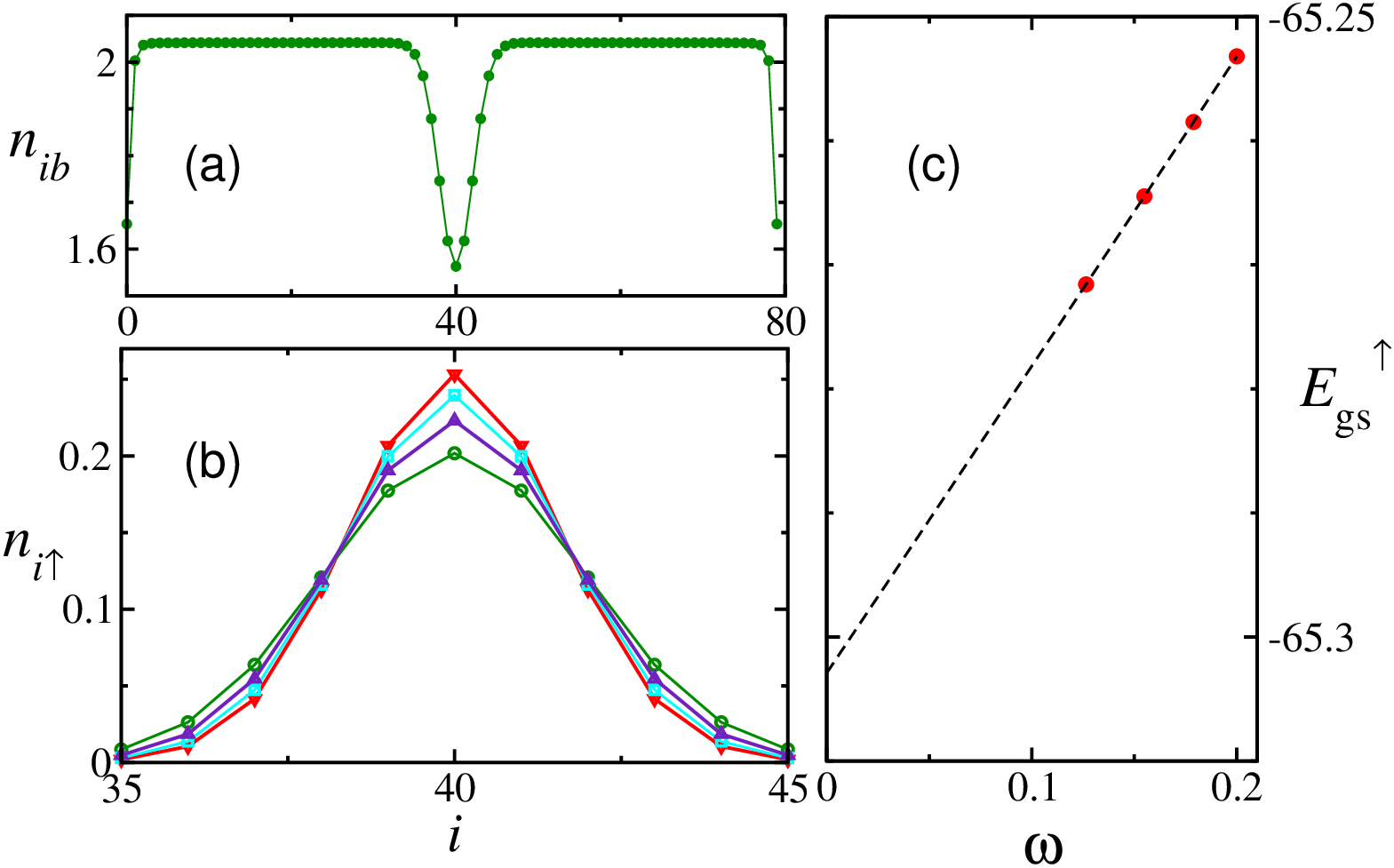}
	\caption{
	A single fermionic impurity immersed in a  one-dimensional Bose gas 
	and trapped at the center of the chain by a shallow harmonic potential of frequency $\omega$. (a) Local density distribution of bosons $n_{ib}$
	for $\omega=0.1265$. (b) Fermionic density profile $n_{i\uparrow}$ near the trap center plotted for different values of the trap frequency 
	$\omega=0.1265$ (circles), $0.1549$ (triangle up), $0.1789$ (square) and $0.2$ (triangle down).
	(c) Ground state energy $E_\mathrm{gs}^{\uparrow}$ of the coupled system as a function of the trap frequency. The dashed line is the quadratic polynomial, $p (\omega)= a_0+ a_1 \omega + a_2 \omega^2$, with fitted parameters $a_0=-65.3029$, $a_1=0.246$ and $a_2=0.0123$.
		The length of the chain is $L=80$ and the boson density is $n_{b} = 2$. Tunneling rates of bosons and fermions are set to $t_b=t_f=1$. 
Boson-boson and boson-fermion interaction strengths are $U_b=2$ and $U_{bf}=6$, respectively. 
	}
     \label{fig:example}
\end{figure}

Let us now consider a finite trap frequency $\omega$ for the impurity and let $E_\mathrm{gs}^{ \uparrow}$ be the corresponding 
 ground state energy of the mixture that we compute by the DMRG method. For small values of $\omega$, it follows  from Eqs.~(\ref{HamF}) and (\ref{poladisp}) 
 that  $E_\mathrm{gs}^{ \uparrow}$ is well approximated by the ground state energy of the following anharmonic quantum oscillator Hamiltonian in first-quantized form:
 \begin{equation}\label{ho}
H_\textrm{an}^\uparrow=E_0^\uparrow+\frac{q^2}{2m_p^*} +\frac{1}{2}m^* \omega^2 x^2 +A^\uparrow q^4,
\end{equation}
where $x=i-L/2$ represents the continuum spatial coordinate of the impurity. 
By treating the anharmonic term in Eq.~(\ref{ho}) within first-order perturbation theory, we obtain
\begin{equation}\label{ana}
E_\mathrm{gs}^{\uparrow}\simeq E_0^\uparrow+\frac{1}{2}\omega_{p} +\frac{3}{4} {m_p^*}^2 A^\uparrow  \omega_{p}^2,
\end{equation}
where $\omega_{p}=\omega \sqrt{m^*/m_{p}^*}$ is the effective trap frequency for the polaron.  
This corresponds to the frequency of the dipole oscillation that can be generated experimentally 
by displacing the trap center of the fermionic impurity, provided that the boson density is kept uniform~\cite{rey2005}. 

Equation (\ref{ana}) suggests that the binding energy and the effective mass of the polaron can be inferred by calculating the ground state energy  $E_\mathrm{gs}^\uparrow $
of the coupled system for several (small) values of the trap frequency $\omega$ and fitting the obtained results 
via a quadratic polynomial
\begin{equation}
\label{eq:pol}
p (\omega)= a_0+ a_1 \omega + a_2 \omega^2,
\end{equation}
 where $a_0$, $a_1$, and $a_2$ are fitting coefficients. By comparing Eq.~(\ref{eq:pol}) with Eq.~(\ref{ana}), we find $E_0^\uparrow=a_0$ and $m^*/m_p^*=4 a_1^2$.  After this trap scaling procedure, the polaron binding energy $\mu_p$ is evaluated from Eq.~(\ref{mu_p}) using the fitted value of $E_0^\uparrow$ as well as the ground state energy of the bosonic bath $E^b_\mathrm{gs}$ (which is independent of $\omega$).
The specific range of  $\omega$ values to be used for the fit depends on the system size $L$ and the boson-fermion coupling strength $U_{bf}$. To avoid finite-size effects, the density distribution of the fermionic impurity must decay sufficiently fast near the edge of the chain, which yields a lower bound on the trap frequency for a given value of $L$ and $U_{bf}$. 

\subsection{Results}
In Fig.~\ref{fig:example}, we show examples of 
the density profiles of the bosonic bath (panel a), and of the fermionic impurity (panel b) calculated for different values of the trap frequency $\omega$. Here $L=80$, $n_b=2$, $U_b=2$ and $U_{bf}=6$.
We see that the boson local density is strongly 
depleted in the center of the chain, where the fermionic impurity is trapped. 
In panel (c) of Fig.~\ref{fig:example}, we display the calculated values of the
ground state energy  of the coupled system  as a function of the trap frequency, together with the  fitted  
polynomial (\ref{eq:pol}). We see that the latter reproduces quite accurately the numerical data  in the low frequency regime. 
The fitting coefficient $a_2$ is typically small (this will no longer be true for bipolarons, see Sec.~\ref{sec:two}).
Taking into account that the calculated ground state energy of the Bose gas alone is $E_\mathrm{gs}^{b}=-70.1557$, we find that the polaron binding energy for the above parameters is  $\mu_p=6.853$, while the obtained value for the inverse effective mass ratio is $m^*/m_p^*=0.242$. 

\begin{figure}
      \includegraphics[width=0.95\columnwidth]{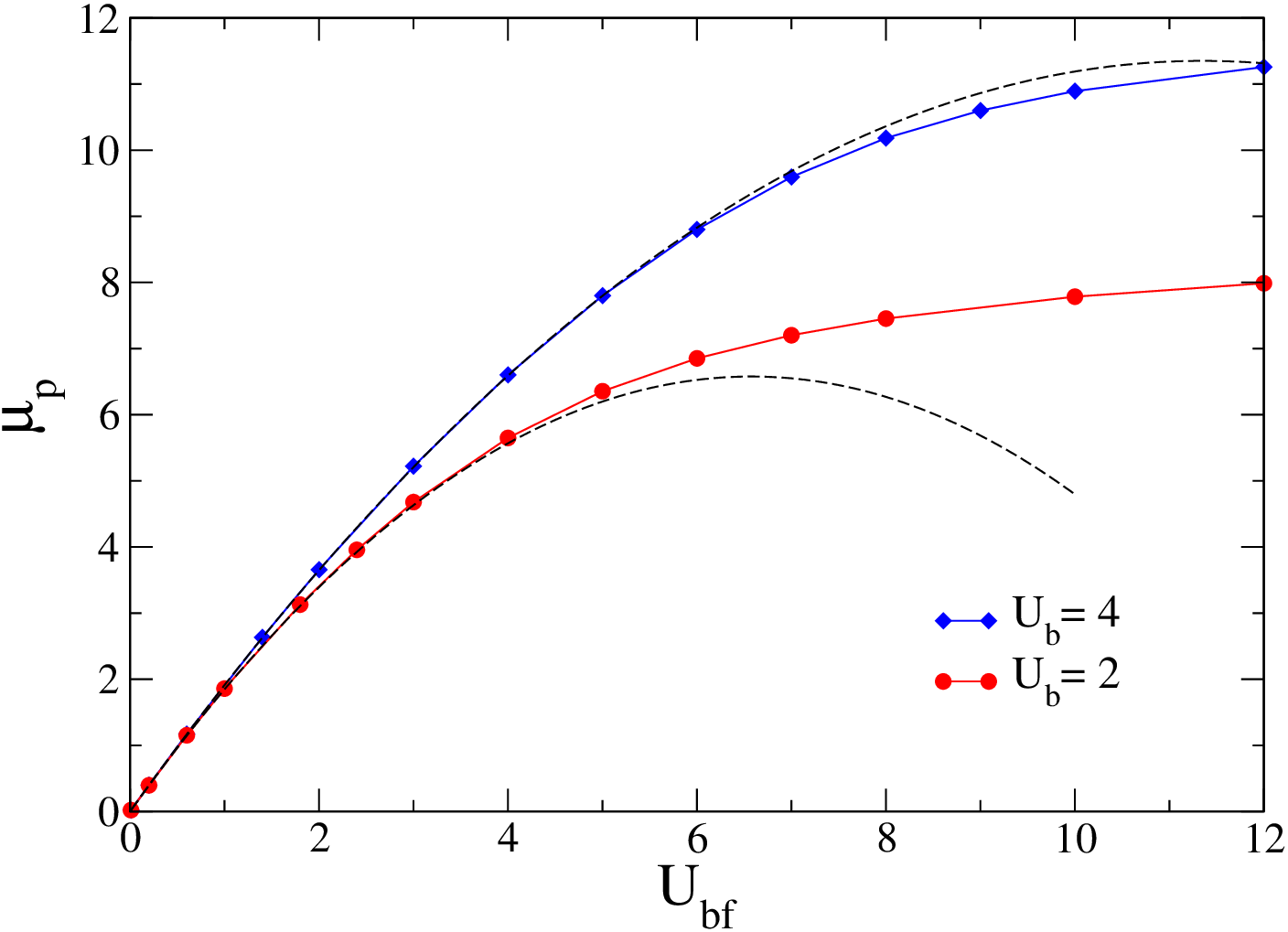}
	\caption{Binding energy $\mu_p$ of the polaron, Eq.~(\ref{mu_p}), as a function of the impurity-bath interaction strength 
	$U_{bf}$, plotted for two different values of the 
	boson-boson repulsion  $U_b=2$ ($\gamma=0.5$, red circles) and $U_b=4$ ($\gamma=1$, blue diamonds). The boson density is set to $n_{b} = 2$.
     Dashed lines correspond to the predictions from Bogoliubov theory, cf.~Eq.~(\ref{eq:bogepol2b}). 
     }
     \label{fig:epol}
\end{figure}

We repeat the above procedure for different values of the boson-fermion coupling $U_{bf}$  and for two  values of the intra-bath interaction strength, $U_b=2$ and $U_b=4$. The obtained results for the polaron binding energy are plotted
 in Fig.~\ref{fig:epol} as a function of $U_{bf}$.  
We see from Fig.~\ref{fig:epol} that the polaron binding energy gets bigger as the coupling to the bath increases, in agreement with previous quantum Monte Carlo studies~\cite{parisi2017,grusdt2017a}.

For weakly-interacting bosons,  the numerical results can be compared against analytical calculations based on Bogoliubov theory
by treating the Bose-Fermi interaction $U_{bf}$ as a small perturbation, as previously done for the continuum~\cite{ardila2015,parisi2017}. 
Up to second-order terms included, one finds  
\begin{equation}
\mu_p^\textrm{bog}=U_{bf} n_{b} - \left( U_{bf} \sqrt{n_{b}}\right)^2 \frac{1}{L} \sum_k \left(u_k + v_k  \right)^2 \frac{1}{  \epsilon_{-k}^f
 -\epsilon_0^f  + \omega_k},
\label{eq:bogepol3}
\end{equation} 
where the first term in the right-hand side corresponds to the mean-field result, while the second term is the correction 
coming from density fluctuations in the bath. The Bogoliubov coefficients $u_k, v_k$ satisfy 
$u_k^2=1+v_k^2=(\overline{\epsilon}_k+U_{b} n_b+\omega_k)/(2 \omega_k)$ and $u_k v_k=-U_{b} n_b/(2 \omega_k)$, 
where $\epsilon^{b}_k=\epsilon^{f}_k$  and  $\omega_k = \left(\overline{\epsilon}_k^2 + \overline{\epsilon}_k 2 n_{b} U_{b} \right)^{1/2}$ is the energy of the elementary excitations with $\overline{\epsilon}_k =2 (1-\cos k)$.
By taking into account that $(u_k+v_k)^2=\overline{\epsilon}_k/\omega_k$ and replacing  the sum over quasi-momenta in Eq.~(\ref{eq:bogepol3}) by an integral, we obtain 
\begin{equation}
\begin{split}
\label{eq:bogepol2b}
       \mu_p^\textrm{bog}=U_{bf} n_{b}     -\frac{U_{bf}^2} {U_b} \left( \frac{1}{2} - \frac{1}{\pi} \arctan \sqrt{\frac{2}{n_{b}U_{b}}} \right).
\end{split}
\end{equation}
In the continuum limit, $n_{b}U_{b} \rightarrow 0$, Eq.~(\ref{eq:bogepol2b}) reduces to 
$\mu_p^\textrm{bog} \simeq n_b U_{bf} - \frac{U_{bf}^2}{\pi} \frac{1}{\sqrt{\gamma}} m^*$, where $m^*$ is the bare mass of the impurity~\cite{parisi2017}. 

We see from Fig.~\ref{fig:epol} that  the Bogoliubov prediction reproduces properly our numerical results for the binding energy in the regime of weak Bose-Fermi coupling, 
but significantly deviates from them in the strong-coupling regime, in particular when the boson-boson interaction strength is small. This is not surprising, as in this limit the second term in the right-hand side of 
Eq.~(\ref{eq:bogepol2b}) becomes dominant over the mean-field correction, and perturbation theory ceases to be valid.

\begin{figure}
       \includegraphics[width=0.95\columnwidth]{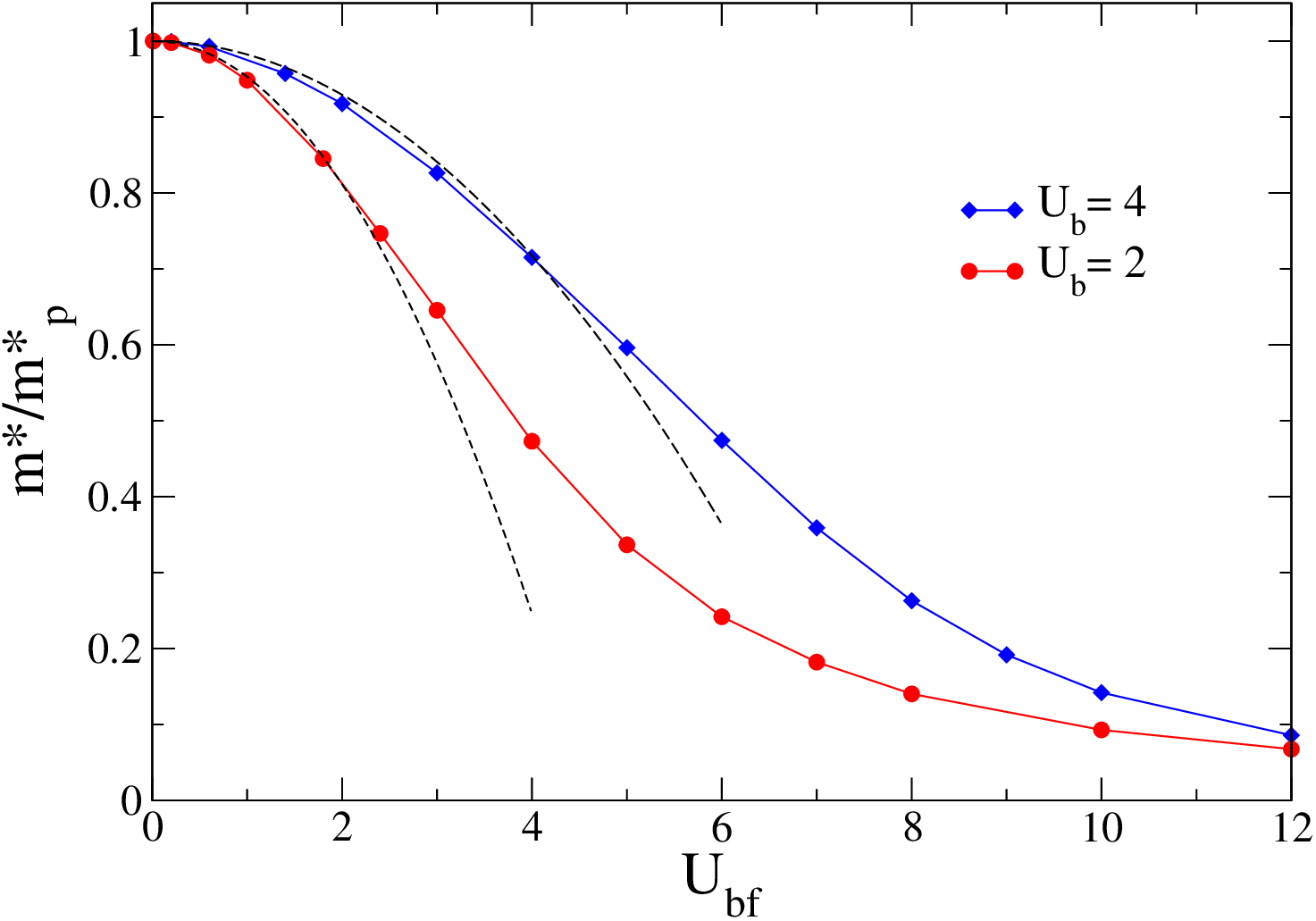}
	\caption{ Inverse effective mass ratio  $m^*/m^*_{p}$  of the polaron as a 
 function of the  impurity-bath interaction strength $U_{bf}$, plotted for two different values of the 
	boson-boson repulsion  $U_b=2$ ($\gamma=0.5$, red circles) and $U_b=4$ ($\gamma=1$, blue diamonds).  The boson density is set to $n_{b} = 2$.
  Dashed lines correspond  to the prediction from Bogoliubov theory, cf.~Eq.~(\ref{effmasspolaron}).}
     \label{fig:effmassp}
\end{figure}

An important effect of the coupling to the bosonic bath is the enhancement of the effective mass  $m_{p}^*$ of the fermionic impurity compared to its bare mass $m^*$, as shown in Fig.~\ref{fig:effmassp}. 
The corresponding Bogoliubov prediction for the inverse of the polaron effective mass can be written as
\begin{equation}\label{bogm}
	\frac{1}{m^{*\mathrm{bog}}_p}=\frac{1}{m^*} +\left. \frac{\partial^2 \mu_p^\mathrm{bog}(q) } { \partial q^2}\right|_{q=0},
\end{equation}
where 
\begin{equation}\label{effmasspolaron0}
 \mu_p^\mathrm{bog}(q) =U_{bf} n_{b} - U_{bf}^2 n_{b} 
 \int_{-\pi}^{\pi} \frac{d k}{2\pi}  (u_k+v_k)^2 
 \frac{1}{\epsilon_{q-k}^f -\epsilon_q^f +\omega_k}
\end{equation}
is the generalization of the Bogoliubov prediction for the binding energy of the impurity, Eq.~(\ref{eq:bogepol2b}), 
under the new assumption that the impurity carries a finite quasi-momentum $q$.  

 We then differentiate twice the right-hand side of Eq.~(\ref{effmasspolaron0}) with respect to $q$ and  calculate analytically  the integral over momentum 
 for $q=0$. From Eq.~(\ref{bogm}) we obtain  
\begin{equation}\label{effmasspolaron}
	\frac{1}{m^{*\mathrm{bog}}_p}=2-U_{bf}^2 \frac{n_b}{2\pi} \frac{4\pi +2\sqrt{2x}(x-2)-8 \arctan\sqrt{\frac{2}{x}} }{x^3},
\end{equation}
with $x=n_{b}U_{b}$. In the continuum limit,  $n_{b}U_{b} \rightarrow 0$, Eq.~(\ref{effmasspolaron}) reduces to $m^{*}/m_p^{*\mathrm{bog}}=1-2 \eta^2/(3 \pi \gamma^{3/2})$, with $\eta=U_{bf}/(2 n_b t_b)$, in agreement with previous work~\cite{parisi2017}. Since there is no mean-field correction for the polaron effective mass, the Bogoliubov prediction works better 
for weak boson-boson interactions, as shown in  Fig.~\ref{fig:effmassp} by the dashed lines.

\section{Two impurities}
\label{sec:two}
We now investigate the formation of bound states of two fermionic impurities, i.e.~bipolarons,  
which are induced solely by the exchange of density fluctuations in the bosonic 
bath~\cite{bijlsma2000,heiselberg2000,efremov2002,viverit2002,matera2003}. 
In this aim, we first generalize the trap scaling procedure introduced in Sec.~\ref{sec:single} to bipolarons. We then use this approach to compute 
the binding energy and effective mass of bipolarons for both the singlet $\uparrow\downarrow$ spin configuration  
(corresponding to $N_\uparrow=N_\downarrow=1$), and the triplet $\uparrow\uparrow$ spin configuration ($N_\uparrow=2$, $N_\downarrow=0$). 

\subsection{Trap scaling approach for bipolarons}
We proceed as in Sec.~\ref{sec:single} by first assuming that there is no trapping potential, i.e.~we set $\omega=0$ in Eq.~(\ref{HamF}), and that the length of the chain is 
infinite. Let $E_Q$ be the lowest energy level of the mixture containing a bipolaron with center-of-mass 
quasi-momentum $Q$. In analogy to Eq.~(\ref{poladisp}), for small $Q$ we can write that
\begin{equation}
\label{poladispQ}
E_Q\simeq E_0+\frac{Q^2}{2M_{bip}^*}+A Q^4,
\end{equation}
where $E_0$ is the ground-state energy of the uniform mixture, $M_{bip}^*$ is the bipolaron effective mass and
 $A$ is a numerical coefficient accounting for anharmonic terms in the dispersion relation. 
The binding energy $E_b$ of the bipolaron is then defined as 
\begin{equation}\label{eq:Eb}
-E_b=E_0-2 E_0^\uparrow + E_\mathrm{gs}^{b},
\end{equation} 
where the minus sign in the left-hand side of Eq.~(\ref{eq:Eb}) ensures that $E_b\geq 0$.

We now discuss the effect of a shallow harmonic trap acting on the two fermionic impurities. Let   $E_\mathrm{gs}$ 
be the corresponding ground state energy of the mixture and let $x_1=i-L/2$, $x_2=j-L/2$ be the spatial coordinates of the two fermions measured with respect to the center of the chain. In first-quantization formalism, 
the total external potential acting on the impurities
 can be written as
\begin{equation} 
\frac{1}{2} m^* \omega^2 (x_1^2+x_2^2)=\frac{1}{2} 2m^* \omega^2 R^2 + \frac{1}{2} \frac{m^*}{2} \omega^2 r^2,
\end{equation}
where $R=(x_1+x_2)/2$ and $r=x_1-x_2$ represent the center-of-mass and the relative motion coordinates, respectively. For bound states, the mean distance between the two
constituent particles is finite, $ \sqrt{\langle r^2 \rangle}<+\infty $, implying that the effect of a shallow trap on the relative motion is perturbative.  
Hence, for a small enough trap frequency, $E_\mathrm{gs}$   is given by 
the ground state energy of the following Hamiltonian:   
\begin{equation}
\label{HanQ}
H_\textrm{an} = E_0 +\frac{1}{4} m^* \omega^2 \langle r^2 \rangle +\frac{Q^2}{2M_{bip}^*}+ \frac{1}{2} 2m^* \omega^2 R^2+ A Q^4,
\end{equation}
where the last term can again be evaluated within first-order perturbation theory. 
This yields
\begin{equation}\label{gsbip}
E_\mathrm{gs}\simeq E_0+ \frac{1}{2} \omega_{bip}+
 \left(\frac{3}{4} {M_{bip}^*}^2 A  + \frac{1}{8}M_{bip}^*  \langle r^2 \rangle\right) \omega_{bip}^2,
\end{equation}
where $\omega_{bip}=\omega \sqrt{2m^*/M_{bip}^*}$  is the effective trap frequency for the bipolaron. 
As previously done for the single polaron problem, we compute the ground state energy of the mixture for different (small) values of the trap 
frequency $\omega$ and fit the obtained results via a quadratic polynomial 
 \begin{equation}\label{tildep}
 \tilde p(\omega)=\alpha_0 +\alpha_1 \omega+\alpha_2 \omega^2,
 \end{equation}
where $\alpha_i$ are fitting parameters. From this, we obtain the ground state energy $E_0=\alpha_0$ of the uniform mixture,
 as well as the inverse effective mass ratio $2 m^*/M_{bip}^*=4\alpha_1^2$ of the bipolaron. The binding energy $E_b$ is then 
 evaluated from Eq.~(\ref{eq:Eb}).
 
 
We emphasize that Eq.~(\ref{gsbip}) relies on the assumption that the effect of the trap on the relative motion of the two impurities is \emph{perturbative} 
and therefore contributes only to  the  $\omega^2$  term  in $E_{gs}$.  
As a consequence, our  fitting procedure is accurate only if  the binding energy of the bipolaron is large compared to the trap 
frequency, $E_b \gg  \omega$.   At the breaking point of the molecule, the average size of the molecule diverges and the rhs of  
Eq.~(\ref{gsbip})  becomes ill defined.
In order to satisfy the  condition $E_b \gg  \omega$, we would need to consider smaller and smaller values of the trap frequency, which in turn require simulations of larger and larger systems. Numerically accurate results in this extreme regime are therefore challenging.

 We have benchmarked the trap scaling procedure 
 for two simpler one-dimensional systems: two fermions with opposite spin obeying the  attractive Fermi-Hubbard model and two identical fermions described by the
 attractive $t-V$ model. In both cases we have recovered the known analytical results~\cite{Valiente_2008} for the bound state \emph{in the absence of the trap}.

\subsection{Results: $\uparrow\downarrow$ bipolaron}
\begin{figure}
     \includegraphics[width=0.95\columnwidth]{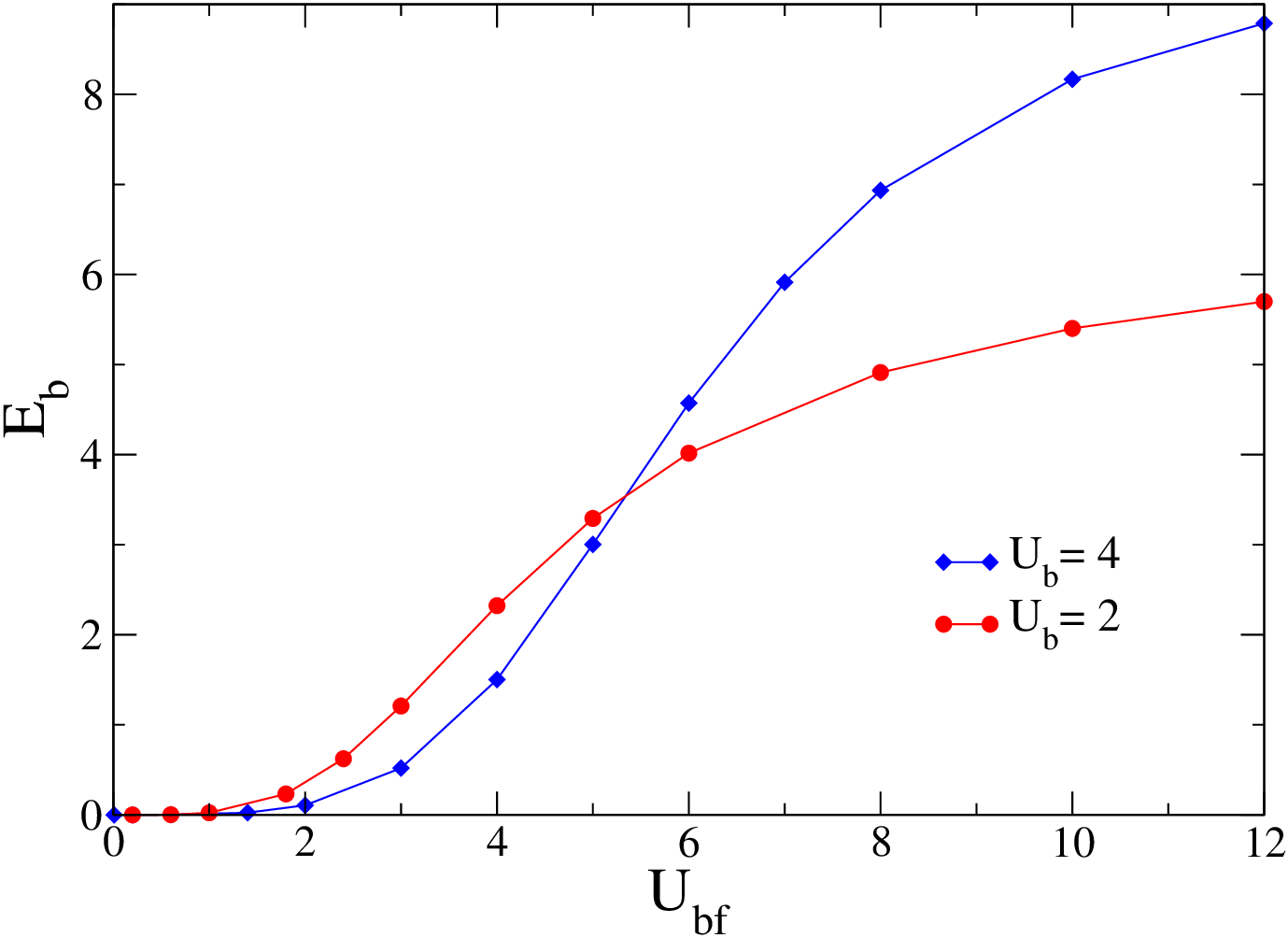}
	\caption{Binding energy $E_{b}$ of the $\uparrow \downarrow$ bipolaron,  cf.~Eq.~(\ref{eq:Eb}), as a function of the impurity-bath interaction strength
	 $U_{bf}$. The two data sets correspond to two different values of the boson-boson interaction, $U_b=2$ (red circles) and $U_b=4$ (blue diamonds). The 
	 boson density is set to $n_b=2$.}
     \label{fig:Eb}
\end{figure}

   In Fig.~\ref{fig:Eb}, we show the numerical results for the binding energy of the $\uparrow\downarrow$ bipolaron as a 
function of the impurity-bath interaction $U_{bf}$ for two different values of the intra-bath interaction, $U_b=2$ and $4$, with the boson density fixed to $n_b=2$. 
The bipolaron binding energy  is found to be always positive for any finite value of the impurity-bath coupling. This is consistent with the fact that in one dimension 
any attractive contact interaction produces a two-body bound state, leading to a BCS instability~\cite{feiguin2012}. 

As shown in Fig.~\ref{fig:Eb}, the binding energy $E_b$ increases monotonously with the impurity-bath coupling $U_{bf}$, and tends to saturate for $U_{bf}\rightarrow +\infty$.
In the regime of weak impurity-bath coupling, we also find that the binding energy of the bipolaron decreases as the boson-boson   
interaction $U_b$ gets larger. This is due to the fact that bosonic density fluctuations, which mediate the effective attractive interaction
between the two impurities, are progressively reduced as the level of correlation in the bath rises. 
Indeed, in the limit $U_b\rightarrow+\infty$, i.e.~deep in the Mott regime for integer fillings~\cite{cazalilla2011}, the bosonic bath acts as a uniform external potential for the fermions and hence the pair binding energy vanishes in the absence of a direct interaction between the two impurities.

In the opposite limit of strong coupling with the bath, the tendency is reversed, namely the binding energy becomes larger for stronger boson-boson interaction,
as can be seen in Fig.~\ref{fig:Eb}. This can be attributed to the fact that for $U_{bf}\gg U_b\gg 1$, the $\downarrow$ fermion
can occupy the same hole as previously created in the bosonic bath to host the  $\uparrow$ fermion without any significant additional energy cost, so that $E_0\simeq E^\uparrow_0$. From Eqs.~(\ref{eq:Eb}) and (\ref{mu_p}), this implies that 
\begin{equation} \label{ex}
E_b\simeq \mu_p-2,
\end{equation}
and thus that $E_b$ should increase with growing $U_b$ (cf.~Fig.~\ref{fig:epol}), 
in full agreement with our numerics. 
 
\begin{figure}
     \includegraphics[width=0.95\columnwidth]{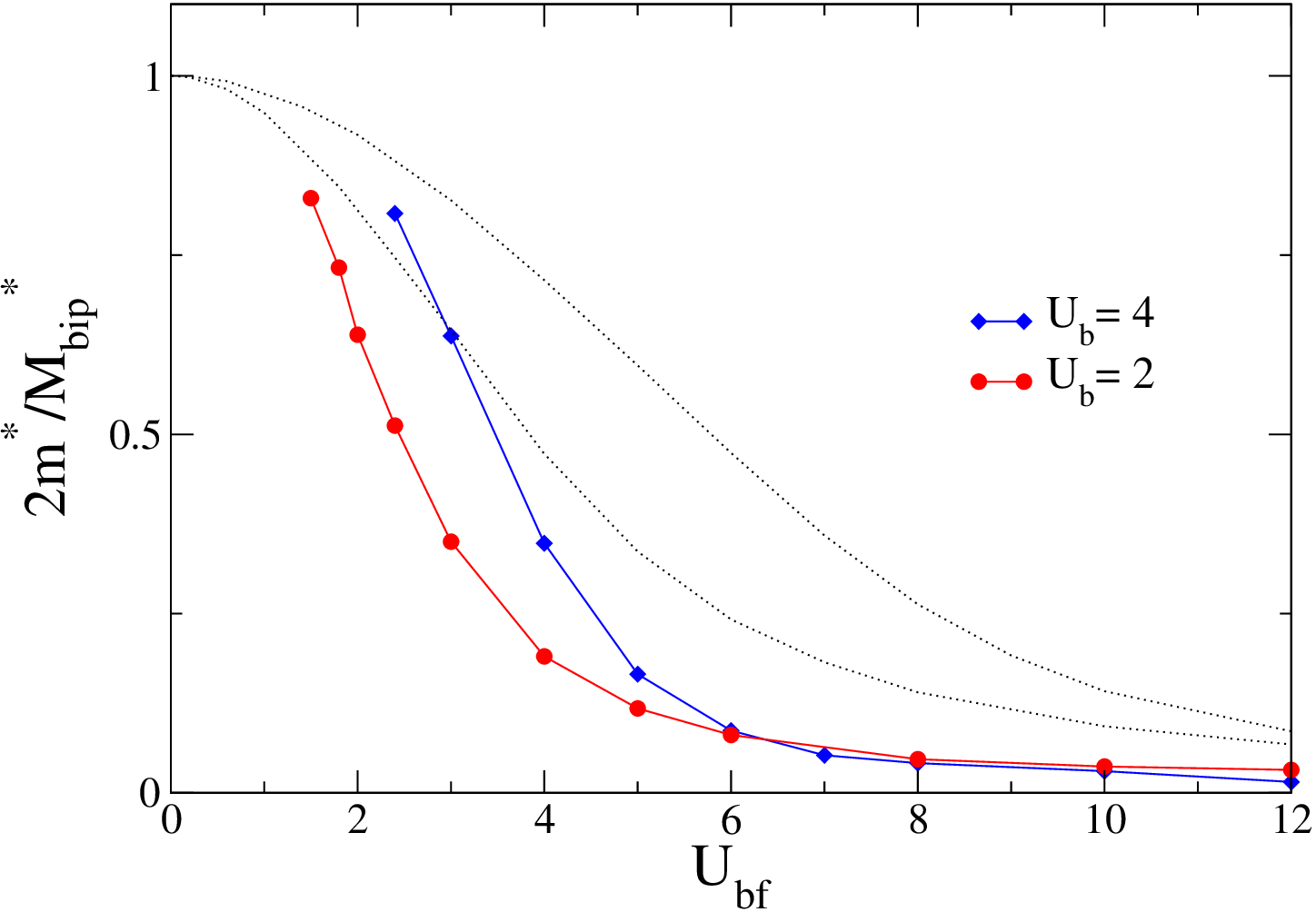}
     \caption{Inverse effective mass ratio of the $\uparrow \downarrow$ bipolaron  as a function of the impurity-bath interaction strength. The two data sets correspond to two different values of the boson-boson interaction, $U_b=2$ (red circles) and $U_b=4$ (blue diamonds). The boson density is set to $n_b=2$. The dotted lines represent the corresponding effective mass ratios for the single polaron, 	 which are reproduced from Fig.~\ref{fig:effmassp}.  For  small values of $U_{bf}$, we do not display any numerical data points as our fitting procedure to extract $M_{bip}^*$ becomes inaccurate due to finite-size effects (see text). In this regime the mass ratios of polaron and bipolaron approach each other. 
	 } 
     \label{fig:m_updown}
\end{figure}

In  Fig.~\ref{fig:m_updown}, we display the corresponding results for the inverse effective mass ratio of the 
$\uparrow\downarrow$ bipolaron, for the same set of system parameters. 
On a general ground, it is expected that the effective mass of the molecule increases as its binding energy builds up (cf.~Fig.~\ref{fig:Eb}). This is indeed what we find, as for weak Bose-Fermi coupling the effective mass $M_{bip}^*$ is reduced for increasing boson-boson repulsion, while in the opposite regime of strong coupling, $U_{bf}\gg 1$, the tendency is reversed. In particular, we see by comparison between  Fig.~\ref{fig:Eb} and \ref{fig:m_updown} that the two curves cross at similar values of the impurity-bath coupling (around $U_{bf}= 6$).

In Fig.~\ref{fig:m_updown}, we do not display results for small values of $U_{bf}$, where the binding energy of the bipolaron (see Fig.~\ref{fig:Eb}) becomes comparable or smaller than the trap frequency, $|E_b|\lesssim \omega$. In this regime the effect of the harmonic trap is not perturbative and our fitting procedure to extract $M_{bip}^*$ is no longer accurate. As this occurs, the bipolaron effective mass $M_{bip}^*$ is already close to the sum of the mass of its constituents $2m_{p}^*$, i.e.~twice the effective mass of a single polaron, as indicated in Fig.~\ref{fig:m_updown} by the dotted lines. 	 

\begin{figure}
     \includegraphics[width=0.95\columnwidth]{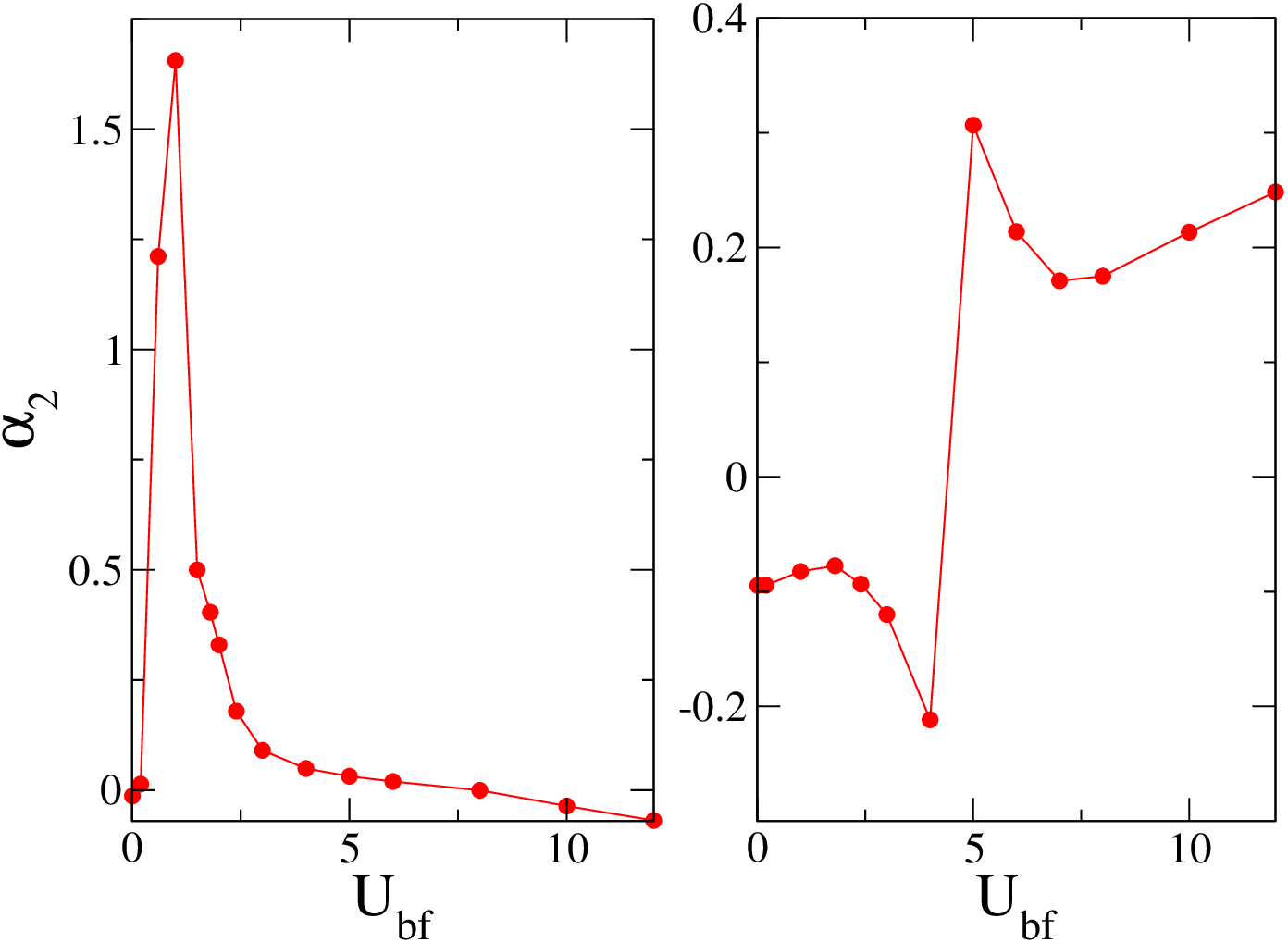}
     \caption{Fitting coefficient $\alpha_2$, cf.~Eq.~(\ref{tildep}), as a function of the Bose-Fermi interaction strength $U_{bf}$ for the $\uparrow\downarrow$ bipolaron (left panel)
     and for the $\uparrow\uparrow$ bipolaron (right panel). The system size is $L=80$. The boson-boson repulsion is set to $U_b=2$, while the boson density is $n_b=2$.}
     \label{fig:alpha2}
\end{figure}

 In the left panel of Fig.~\ref{fig:alpha2}, we show the dependence of the fitting coefficient $\alpha_2$ on the coupling strength $U_{bf}$, for the case $U_b=2$ ($U_b=4$ yields similar results).  
 Due to the finite harmonic oscillator length of the trap, we see that $\alpha_2$ does not diverge as $U_{bf}$ approaches zero as expected from the divergence of the bound state size $\langle r^2\rangle$, but reaches a maximum and then
decreases until it turns negative. This stems from the fact that at $U_{bf}=0$, where no bound state exists, one is left with the problem of two non-interacting lattice fermions
in a harmonic potential. 
 In a previous work~\cite{chalbaud1986}, an analytical formula was found for the perturbative expansion of the single-particle energy levels of the quantum harmonic oscillator on a lattice as a function of the trap frequency $\omega$. For two fermions with opposite spins we find that  $\alpha_1=1$ and $\alpha_2=-1/32$, whereas for equal spins  $\alpha_1=2$ and $\alpha_2=-3/32$ (in both cases $\alpha_0=-4$). 
These limiting results are in full agreement with our numerical results displayed in Fig.~\ref{fig:alpha2}.

\subsection{Results: $\uparrow\uparrow$ bipolaron}
Let us now consider the case of two fermionic impurities with the same spin state. Despite the Pauli exclusion principle which forbids any contact interaction between the two impurities, the formation of bipolarons is nevertheless possible owing to the non-local nature of the phonon-mediated interactions~\cite{recati2005,waechter2007,schecter2014,pavlov2018,reichert2018,reichert2019}, although the typical values of their pair binding energy are smaller compared to the spin singlet configuration.

\begin{figure}
     \includegraphics[width=0.95\columnwidth]{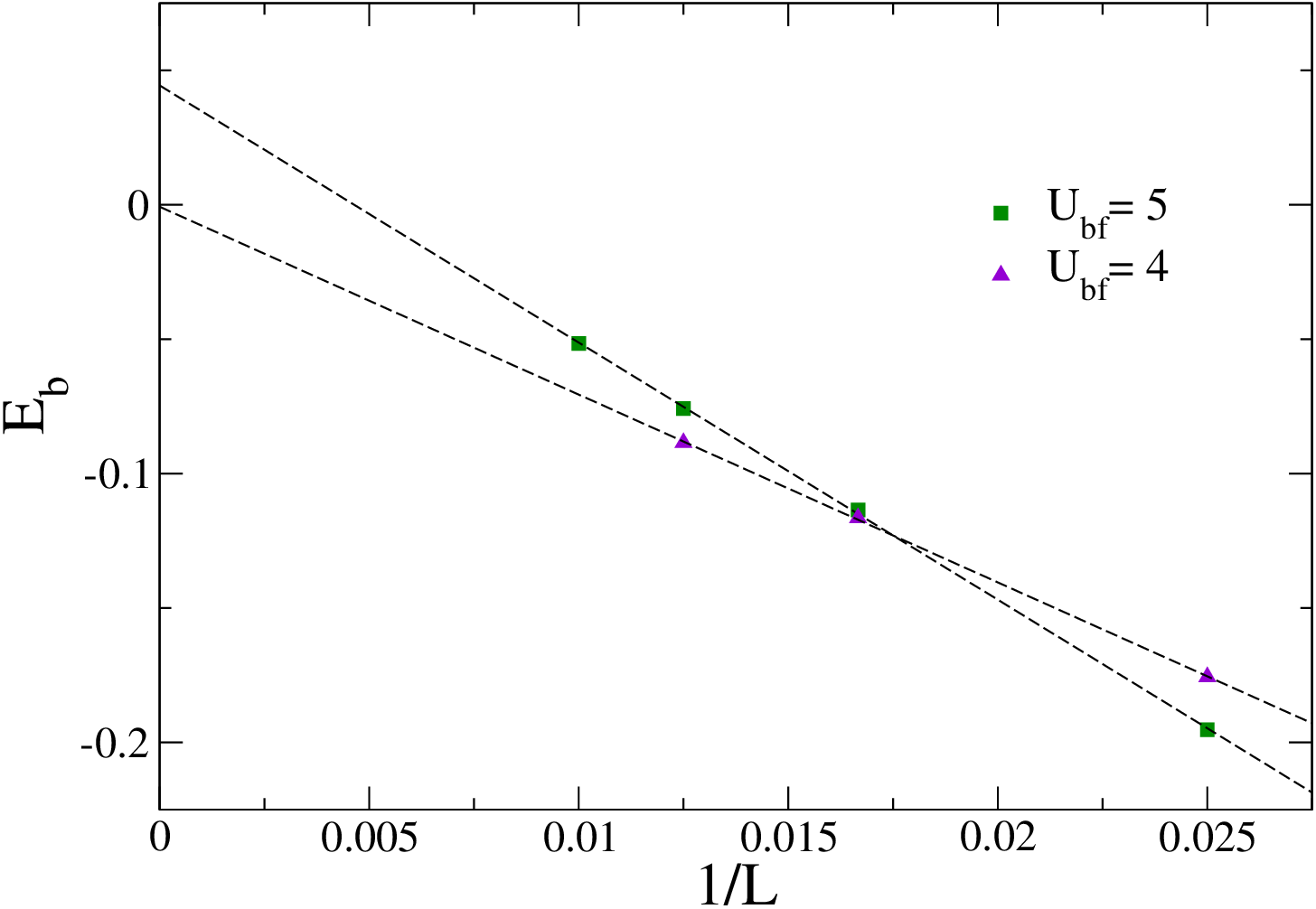}
	\caption{Binding energy of the $\uparrow\uparrow$ bipolaron as a function of the inverse system size $1/L$ for two different values of the Bose-Fermi coupling, $U_{bf}=4$ (triangles) and $U_{bf}= 5$ (squares). The bosonic bath parameters are $n_b=2$ and $U_b=2$, and the length of the chain varies from $L=40$ to $L=100$. The numerical data are well 
	fitted with a straight line, whose $y$-intercept yields an estimate of $E_b$ in the infinite system. 
	For $U_{bf}=4$ the $y$-intercept vanishes, signaling the absence of a bound state at low $U_{bf}$.}
     \label{fig:scalingEbuu}
\end{figure}

Importantly, we find that for the $\uparrow\uparrow$ bipolaron formation a finite size scaling analysis is crucial to obtain meaningful results. 
 As an example, in Fig.~\ref{fig:scalingEbuu} we plot the values of the pair binding energy, calculated via the trap scaling procedure,
as a function of the inverse size of the chain $1/L$ for two different values of the Bose-Fermi coupling.
All the obtained values of the pair binding energy shown in Fig.~\ref{fig:scalingEbuu} are negative, indicating the presence
of strong finite-size effects. 
However, both data sets show a linear dependence in $1/L$, and the intercept of the curves with the $y$ axis yields the extrapolated value of the pair binding energy 
in the thermodynamic limit, $L\rightarrow +\infty$.
For $U_{bf}=4$, the $y$-intercept is vanishingly small, signaling the absence of bipolarons, whereas for $U_{bf}=5$ the 
$y$-intercept is finite and positive, implying that a bipolaron has formed in the system. 
 Hence, we find that impurities with equal spins form a spin-triplet bound state only above a finite critical interaction strength $U_{bf}^c$, in sharp 
 contrast with the singlet case. 
As shown in the right panel of Fig.~\ref{fig:alpha2}, the value of the coefficient $\alpha_2$ changes from negative to positive 
in the vicinity of the critical point, providing  a simple way to approximately estimate its position.

\begin{figure}
     \includegraphics[width=0.95\columnwidth]{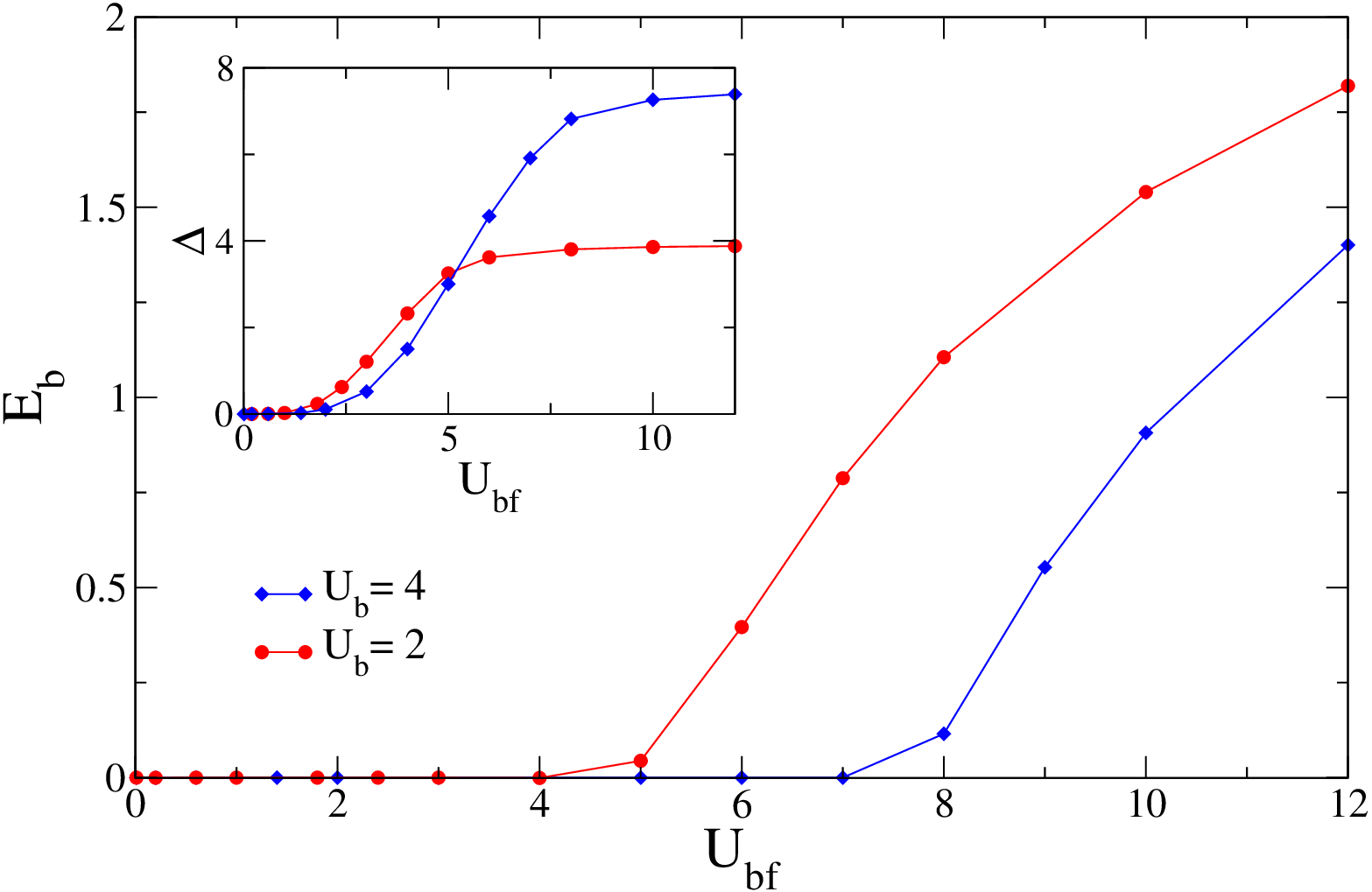}
	\caption{Main panel: Pair binding energy $E_b$  of the $\uparrow\uparrow$ bipolaron, extrapolated to infinite system size (see Fig.~\ref{fig:scalingEbuu}), as a function of the impurity-bath interaction,
	for two values of the boson-boson repulsion $U_b=2$ (red circles) and $U_b=4$ (blue diamonds). The boson density is $n_b=2$.
	The  bipolaron is formed only beyond a finite critical value  of the impurity-bath coupling $U_{bf}^c$. Inset: spin energy gap $\Delta$ of
	bipolarons,  cf.~Eq.~(\ref{eq:Delta}), versus Bose-Fermi coupling.}
     \label{fig:Ebuu}
\end{figure}

In Fig.~\ref{fig:Ebuu}, we display the pair binding energy for the $\uparrow\uparrow$ triplet bipolaron state as a function of the Bose-Fermi coupling strength using the same set of parameters as for the single polaron case. We see that mixtures with strongly-correlated bosons require a larger value of the critical Bose-Fermi interaction $U_{bf}^c$ for the formation of triplet polaron pairs to happen, in agreement with early bosonization studies~\cite{mathey2004}.
A recent theoretical study on fermionic impurities in three-dimensional Bose-Einstein condensates found a similar dependence of the $p$-wave critical interaction strength 
on the bath interaction parameter~\cite{camacho2018b}.
We see from Fig.~\ref{fig:Ebuu} that, for a fixed value of $U_{bf}$, the pair binding energy is inversely proportional to the intra-bath interaction strength $U_b$.
Unlike that for the $\uparrow \downarrow$ singlet bipolaron, this behavior extends in the strong-coupling regime 
of large $U_{bf}$, where the pair binding energy saturates to $E_b\simeq 2.34$ and $E_b\simeq 1.89$ for $U_b=2$ and $U_b=4$, 
respectively (not shown in Fig.~\ref{fig:Ebuu}).
 Since the Pauli exclusion principle prevents the two fermionic impurities with equal spin 
from sharing the same site, two different holes in the bosonic bath have to be created at neighboring sites to accommodate the impurities. As a consequence, the simple relation between the binding energies of the polaron and the bipolaron in Eq.~(\ref{ex}) does not hold for the $\uparrow \uparrow$ bipolaron. 

It is also interesting to investigate the spin gap associated to the bipolaron state, which is defined as the difference between
the pair binding energies of the singlet and triplet spin configurations: 
\begin{equation}
\label{eq:Delta}
\Delta=E_b^{\uparrow\downarrow}-E_b^{\uparrow\uparrow}.
\end{equation}
Our numerical results for the spin gap are displayed in the inset of Fig.~\ref{fig:Ebuu}. We see that $\Delta$ becomes nearly constant once
the ${\uparrow\uparrow}$ bipolaron formation sets in for $U_{bf}>U_{bf}^c$.

 \begin{figure}
     \includegraphics[width=0.95\columnwidth]{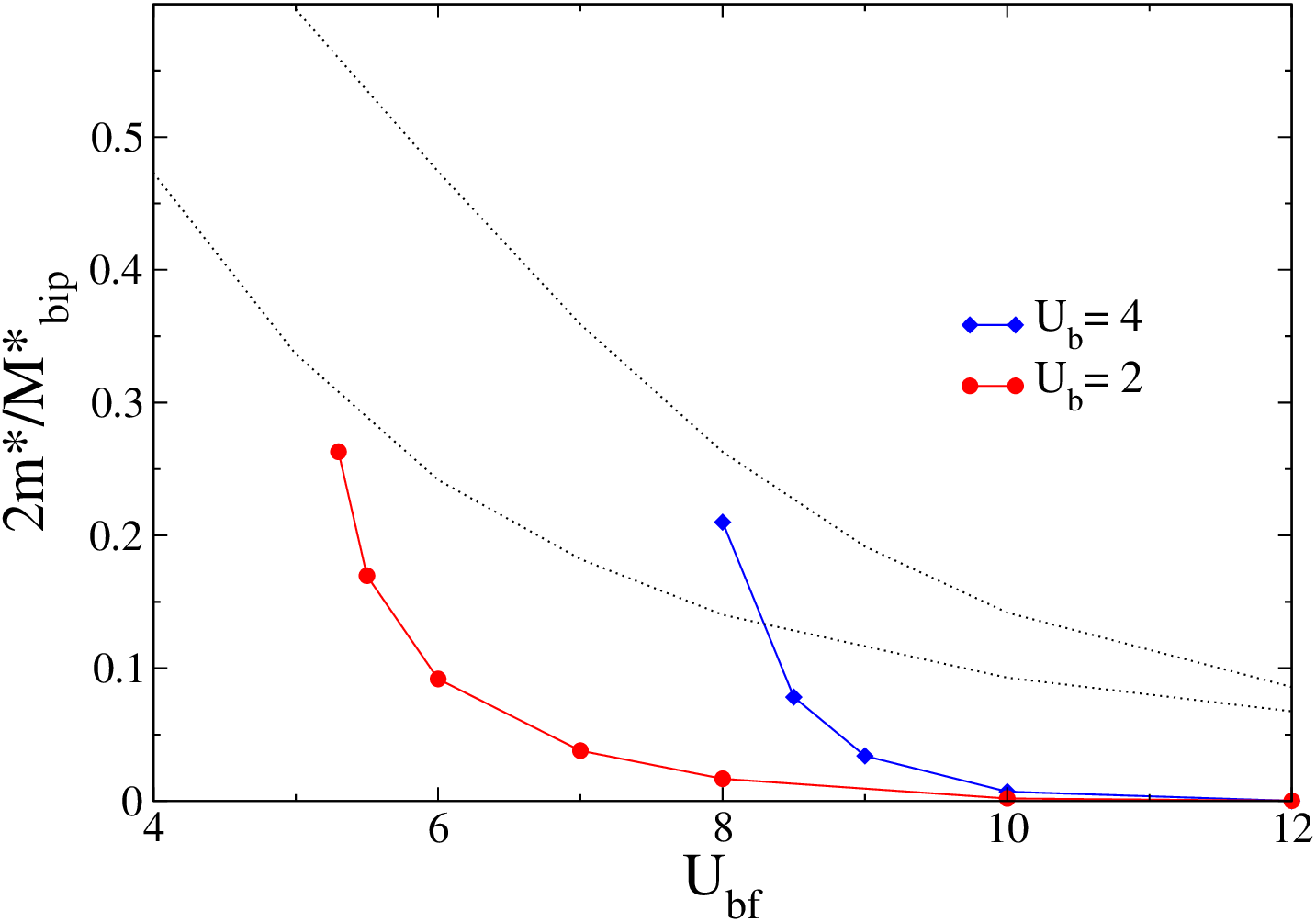}
	\caption{Inverse effective mass ratio of the $\uparrow\uparrow$ bipolaron as a function of the impurity-bath interaction strength
	 $U_{bf}$. The two data sets correspond to two different values of the boson-boson interaction, $U_b=2$ (red circles) and $U_b=4$ (blue diamonds). The 
	 boson density is set to $n_b=2$. The dotted lines represent the corresponding effective mass ratios for the single polaron, 
	 reproduced from Fig.~\ref{fig:effmassp}. Close to the breaking point of the bipolaron state (see Fig.\ref{fig:Ebuu}), where 
	 the mass ratios of polaron and bipolaron approach each other, our numerical data are biased by finite-size effects and are therefore not shown (see text).  
	 }
     \label{fig:Muu}
\end{figure}

In Fig.~\ref{fig:Muu}, we show the corresponding results for the effective mass ratio $2m^*/M^*_{bip}$ of the ${\uparrow\uparrow}$ bipolaron as a function of the Bose-Fermi coupling.   
Near the breaking point of the bound state, $U_{bf}=U_{bf}^c$, we find that the effective mass ratio of the bipolaron approaches that of the single polaron, $m^*/m^*_{p}$. 
By comparing Fig.~\ref{fig:Muu} with Fig.~\ref{fig:m_updown}, we see that the effective mass of the bipolaron  in the triplet state 
increases much faster with the Bose-Fermi coupling than the effective mass of the singlet state. This interesting effect can be understood from the fact that the motion of the $\uparrow\uparrow$ bipolaron requires a significant rearrangement of the bosonic bath in order to shift the two neighboring holes hosting the bound state, resulting in an increased inertia.

\section{Conclusions}
\label{sec:conc}
To summarize, we have presented a thorough numerical study of a few noninteracting spin$-1/2$ fermions coupled to a one-dimensional 
gas of correlated lattice bosons. 
We found that despite the absence of direct attractive interaction between the  impurities, the latter can bind in pairs through the exchange of density fluctuations in the bosonic bath. In order to fully characterize the ground-state properties of this system, the binding energy and the effective mass of both polarons and bipolarons have been investigated numerically through accurate DMRG calculations, based on a novel trap scaling procedure. 

For the bipolaron state, we have shown that the binding energy and effective mass exhibit qualitatively different behavior depending on the  spin state of the two impurities. For opposite spin states, a bipolaron bound state exists for any finite value of the impurity-bath (or Bose-Fermi) coupling, while for impurities with equal spin states, bipolarons appear only beyond a critical value of the coupling strength to the bath. 
In the $\uparrow\downarrow$ (singlet) spin configuration, 
the dependence of the binding energy and effective mass of the bipolaron on the  bath interaction parameter  
exhibits an opposite trend in the weak and in the strong coupling regimes, a feature not present in the $\uparrow\uparrow$ triplet spin configuration. 
Furthermore, we found that for equal spin states the effective mass of the bipolaron, once formed, grows much faster as the strong coupling regime is approached than in the singlet spin configuration.


Our results can be tested in future experiments on bipolarons in one-dimensional Bose-Fermi mixtures of ultra-cold atoms. We believe that the whole parameter space that we studied here should be accessible to experiments, as the impurity-bath Bose-Fermi coupling can be tuned via a Feshbach resonance, while the interaction parameter of the bosonic bath can be controlled
by changing the depth of the optical lattice in the longitudinal direction. 

\begin{acknowledgments}
We acknowledge fruitful discussions with F.~Chevy, C.~Salomon, F.~Werner, S.~Giorgini, J.H.~Thywissen and F.~Grusdt. This
work was supported by ANR (grant SpifBox).
\end{acknowledgments}

\bibliography{cooper_polaron}

\end{document}